\documentclass[%
 reprint,
 amsmath,amssymb,
pra,
]{revtex4-2}

\usepackage{graphicx}
\usepackage{dcolumn}
\usepackage{bm}

\usepackage{blindtext}

\begin{document}


\title{Itinerant ferromagnetism in dilute SU(N) Fermi gases}
\author{Jordi Pera}
\author{Joaquim Casulleras}
\author{Jordi Boronat}%
\affiliation{%
 Departament de F\'\i sica, Campus Nord B4-B5, Universitat Polit\`ecnica de 
Catalunya, E-08034 Barcelona, Spain
}%

\date{\today}

\begin{abstract}
We present exact analytic results for the energy of a SU(N) repulsive Fermi gas as a
function of the spin-channel occupation at second order in the gas parameter. 
This is an extension of previous results that now incorporates the degree of 
polarization of the system. Therefore, the magnetic properties of the gas can 
be obtained, free from numerical uncertainties. For spin 1/2 we find that 
second-order corrections change the itinerant ferromagnetic transition from 
continuous to first-order. Instead, for spin larger than 1/2 the phase 
transition is always of first-order type. The transition critical density 
reduces when the spin increases, making the phase transition more accessible to 
experiments with ultracold dilute Fermi gases. Estimations for Fermi gases of Yb 
and Sr with spin 5/2 and 9/2, respectively,  are reported.
\end{abstract}

\maketitle


\section{Introduction}
At low density an electron gas is paramagnetic rather than ferromagnetic because 
its energy is essentially kinetic. At a same density, the Fermi energy of a 
polarized (ferromagnetic) phase is larger than the one of the unpolarized 
(paramagnetic) one~\cite{vignale}. However, increasing the density produces an 
increase of the 
potential energy due to the interaction of electrons with different spin. At 
some point, this increment of potential energy can exceed the kinetic 
energy gap and the gas will become ferromagnetic. This is the well-known 
Stoner scenario of itinerant ferromagnetism~\cite{stoner} where a quantum 
magnetic transition is predicted in the absence of any crystal pattern and any magnetic field. 

However, the observation of itinerant ferromagnetism in different materials has 
been extremely elusive because an increase in the density can produce non-Fermi 
liquids~\cite{pfleiderer} or crystallization~\cite{leduc} before the expected 
ferromagnetic transition.
The high tunability of trapped cold Fermi gases has offered a new platform to 
study itinerant ferromagnetism. However, even in this case, the observation of 
this transition has turned elusive because the repulsive branch is 
metastable with respect to the formation of spin up-spin down dimers. A first 
pioneering observation of the ferromagnetic transition \cite{jo} was 
later   revised and concluded that pair formation precludes the achievement 
of the relatively large gas parameters required for observing this transition 
\cite{sanner}. More recently, it has been claimed that the ferromagnetic state 
is effectively obtained after the observation of  spin-domain inmiscibility in 
Fermi $^6$Li around a gas parameter $x=k_F a_0\simeq 1$, with $k_F$ the Fermi 
momentum and $a_0$ the s-wave scattering length \cite{valtolina}. The 
theoretical study of repulsive Fermi gases, to determine the gas parameter at 
which the ferromagnetic transition appears, has been intense 
\cite{conduit,pilati,chang,conduit2,massignan,cui,arias,pilati2}. Itinerant 
ferromagnetism is predicted to happen around $ x \simeq 1$, with the most 
reliable results derived from quantum Monte Carlo (QMC) simulations 
\cite{pilati}. At this 
value of the gas parameter, the Stoner model is not quantitatively accurate and 
the use of microscopic approaches as QMC is the best tool in spite of the sign 
problem.

The Stoner model is derived using first-order perturbation theory on the gas 
parameter, that is it is the Hartree-Fock model for the repulsive Fermi 
gas~\cite{stoner}. 
For particles with spin $1/2$, this model predicts a continuous transition from 
a paramagnetic phase to a ferromagnetic one at $x=\pi/2$. The transition point 
obtained with QMC is significantly smaller~\cite{pilati}, pointing to the 
relevance of 
interactions which Stoner model incorporates only at first order. The first 
correction to the Hartree-Fock energy was obtained by Huang, Lee, and Yang 
using pseudopotentials for the hard-sphere Fermi gas~\cite{huangyang,leeyang}. 
The same result was also 
derived by Galitskii~\cite{galitskii}, using Green functions, and   Abrikosov 
and Khalatnikov by 
means of Landau theory of Fermi liquids~\cite{abrikosov}. This analytical 
result for the 
second-order perturbation theory gives the total energy of the Fermi gas as a 
single function of the gas parameter. However, one cannot use this equation 
when the number of particles in each spin is different or, in other words, the 
polarization does not enter in it as a variable.    

In recent years, the experimental production of SU(N) fermions has renewed the theoretical interest in their study.   Cazalilla \textit{et al.}~\cite{cazalilla_1}  discussed that Fermi gases made of alkaline atoms with two electrons in the external shell, such as $^{173}$Yb, present an SU(N) emergent symmetry. They also claimed that for $s>1/2$ the ferromagnetic transition had to be of first order because the mathematical structure of SU(N$>$2) is significantly different than the one of SU(2). In 2014,  Cazalilla and Rey~\cite{cazalilla_2} reviewed the progress made with ultra-cold alkaline-earth Fermi gases. Ref.~\cite{first_observation} was one of the first observations of SU(N) symmetric interactions in alkaline-earth-metal atoms. Interaction effects in SU(N) Fermi gases  as a function of N were studied in Ref.~\cite{effect_vary_N}: for weak interactions it was predicted that  inter-particle collisions are enhanced by N. Collective excitations in SU(N) Fermi gases with tunable spin 
proved to be instrumental to investigate collective properties of large spin systems~\cite{collective_excitations}. On the other hand, Ref. \cite{prethermalization} studied the prethermalization of these systems starting from different initial conditions finding that, under some conditions, the imbalanced initial state could be stabilized for a certain time. Recently, Ref.~\cite{estronci_N_10} performed a thorough study of the thermodynamics of deeply degenerate SU(N) Fermi gases using $^{87}$Sr for which N can be tuned up to 10.  For temperatures above the super-exchange energy, the behavior of the thermodynamic quantities was found to be universal with respect to N~\cite{universal_behavior}.

In the present work, we solve analytically the second-order perturbation energy 
including the polarization or, more precisely, the dependence on the relative 
concentration of particles with different $z$-spin component. Our results are 
derived for a generic spin and thus, we can apply it to SU(N) fermions. This 
generalization allows, for instance, the study of dilute Fermi gases of 
Ytterbium~\cite{pagano}, with spin 5/2, and Strontium~\cite{goban}, with spin 
9/2, that have been already 
produced in experiments. With the analytic result for the energy, as a function 
of the gas parameter and occupation of spin channels, we perform a Landau 
analysis of the ferromagnetic transition for any spin. We find that, for spin 
1/2, the phase transition turns to be first-order with respect to the 
polarization instead of the continuous character of it derived in the Stoner 
model. Interestingly, the critical density for itinerant ferromagnetism is 
observed to decrease monotonously when the spin increases, opening new 
possibilities for experimental realizations in cold Fermi gases.

\section{Methodology}
We study a repulsive Fermi gas at zero temperature with spin $S$ and spin  
degeneracy $\nu=2 S +1$. In the dilute gas regime, only particles 
with different $z$-spin component interact via a central potential 
$V(r)$ ($s$-wave scattering). The number of particles in each spin channel is 
$N_\lambda=C_\lambda N/\nu$, with $N$ the total number of particles and   
$C_\lambda$ being the fraction of $\lambda$ particles (normalized to be one if the 
system is unpolarized, $N_\lambda=N/\nu$, $\forall \lambda$).  The Fermi  
momentum of each species is $k_{F,\lambda}=k_F C_\lambda^{1/3}$, with $k_F=(6\pi^2n/\nu)^{1/3}$. The kinetic 
energy is readily obtained,
\begin{equation}
\frac{T}{N}=\frac{3}{5}\epsilon_F\frac{1}{\nu}\sum_{\lambda}C_{\lambda}^{5/3},
\label{ekin}
\end{equation}
with $\epsilon_F = \hbar^2k_F^2/(2m)$ the Fermi energy. The lowest-order contributions to the 
potential energy are given by~\cite{bishop}
\begin{equation}
\begin{aligned}
V=\frac{\hbar^2 
\Omega}{2m}\sum_{\lambda_1,\lambda_2}\int\frac{d\textbf{l}}{(2\pi)^3} 
n_l\int\frac{d\textbf{k}}{(2\pi)^3}n_k
     \hspace{2cm}
     \\ 
\times \{ K(\textbf{k},\textbf{l};\textbf{k},\textbf{l})-\delta_{\lambda_1,
\lambda_2 } K(\textbf{k} , \textbf{l};\textbf{l} , \textbf{k}) \} \ ,
\end{aligned}
\label{kmatrix}
\end{equation}
with $\Omega$ the volume, $n_l$ and $n_k$ the momentum distributions of the 
free Fermi gas. Up to second order in the $s$-wave scattering length $a_0$, the 
scattering $K$ matrix is given by~\cite{bishop,baker}   
\begin{equation} 
  K(r,R)=4\pi a_0+(4\pi a_0)^2 \, I(r,R)+O(a_0^3) \ ,
  \label{k2}
\end{equation}
with $\textbf{r}=(\textbf{k}-\textbf{l})/2$ and 
$\textbf{R}=\textbf{k}+\textbf{l}$, the relative and total momentum in the 
center of mass frame, respectively. The function $I(r,R)$ is defined as
\begin{eqnarray}
\lefteqn{ I(r,R)  = \frac{1}{(2\pi)^3}}   \\
       & &     \times \int 2 \, d\textbf{q}d\textbf{q}'\frac{1-(1-n_q)(1-n_{q'})}{q^2+q'^2-k^2-l^2}\delta(\textbf{q}+\textbf{q}'-\textbf{k}-\textbf{l})  \ . \nonumber
\label{Iint}          
\end{eqnarray}
Considering only the first term in the expansion of the $K$-matrix (\ref{k2}), 
one gets for the potential energy (\ref{kmatrix}) the well-know Hartree-Fock 
energy~\cite{stoner},
\begin{equation}
 \left( \frac{V}{N} \right)_1=   \frac{2 \epsilon_F}{3\pi}  \left[ \frac{1}{\nu}\sum_{\lambda_1,\lambda_2}
 C_{\lambda_1}C_{\lambda_2} (1-\delta_{\lambda_1,\lambda_2}) \right] \, x \ ,
 \label{hfenergy}
\end{equation}
with $x \equiv k_F a_0$ the gas parameter of the Fermi gas.

The second order term in the gas parameter $x$ is due to  the second term of the $K$-matrix. This second order term reads
\begin{equation}
\left( \frac{V}{N} \right)_2= \frac{\epsilon_F}{k_F^7} \left[ \frac{1}{\nu}\sum_{\lambda_1,\lambda_2}I_2(k_{F,\lambda_1},k_{F,\lambda_2})(1-\delta_{\lambda_1,\lambda_2}) \right] x^2 \ ,
\label{2ndorder}
\end{equation}
with
\begin{equation}
\begin{aligned}
I_2(C_{\lambda_1},C_{\lambda_2})=\frac{3}{16\pi^5}\int d\textbf{l} \, n_l\int 
d\textbf{k} \, n_k \int 2 \, d\textbf{q}d\textbf{q}'\\
     \times\frac{1-(1-n_q)(1-n_{q'})}{q^2+q'^2-k^2-l^2}\delta(\textbf{q}+\textbf{q}'-\textbf{k}-\textbf{l}) \ .
\end{aligned}
\label{i2tot}
\end{equation}

The calculation of $I_2(C_{\lambda_1},C_{\lambda_2})$ (\ref{i2tot}) is 
rather involved. For the unpolarized phase, i.e., when all the spin states are 
equally populated, the integral in Eq. (\ref{i2tot}) was made in the fifties 
of the past century~\cite{leeyang,huangyang,galitskii,abrikosov}. On the other 
hand, when the gas has a finite polarization 
that integral becomes more cumbersome. In previous work, it was solved 
partially but with a final numerical integration~\cite{piotr}. We have been 
able to 
integrate Eq. (\ref{i2tot}) and found an analytical expression for it (See App. \ref{appendix:c}). Our 
result is the following,
\begin{equation}
I_2(C_{\lambda_1},C_{\lambda_2})  
=\frac{4k_F^7}{35\pi^2}C_{\lambda_1}C_{\lambda_2}\frac{C_{\lambda_1}^{1/3} +C_ 
{\lambda_2}^{1/3}}{2} \, F(y) \ ,
\label{i2final}
\end{equation}
with
\begin{widetext}
\begin{equation}
F(y)= \frac{1}{4}\big(15y^2-19y+52-19y^{-1}+15y^{-2}\big)   
+\frac{7}{8}y^{-2}\big(y-1\big)^4\big(y+3+y^{-1}\big)\ln{\bigg\vert\frac{1-y}{
1+y}\bigg\vert}
    -\frac{2y^4}{1+y}\ln{\bigg\vert 
1+\frac{1}{y}\bigg\vert}-\frac{2y^{-4}}{1+y^{-1}}\ln{\bigg\vert 1+y\bigg\vert}
\label{funy}
\end{equation}
\end{widetext}
and $y\equiv (C_{\lambda_1}/C_{\lambda_2})^{1/3}$. Our result reproduces the 
formula 
derived by Kanno~\cite{Kanno} for the specific case of $s=1/2$ and hard spheres.

Assembling it all together, we can write the energy per particle up to second order 
in $x$, and for any occupation of the $\nu$ available spin states, as
\begin{eqnarray}
\frac{E}{N} & =  & \frac{3 \epsilon_F}{5 \nu} \left\{ 
\sum_{\lambda}C_{\lambda}^{5/3} + \frac{10}{9 \pi}  
\left[ \sum_{\lambda_1,\lambda_2}
 C_{\lambda_1}C_{\lambda_2} (1-\delta_{\lambda_1,\lambda_2}) \right] \, x  
\right.  \nonumber \\
 & + & \left. \frac{5}{3 k_F^7} 
\left[ \sum_{\lambda_1,\lambda_2} I_2(C_{\lambda_1},C_{\lambda_2})(1-\delta_{
\lambda_1,\lambda_2}) \right] \, x^2 \right\} \ ,
\label{ener2}
\end{eqnarray}
with $I_2(C_{\lambda_1},C_{\lambda_2})$ given by Eqs. 
(\ref{i2final},\ref{funy}). Eq. (\ref{ener2}) is a perturbative expansion in $x$ and works fine for low values of $x$,  $x<1$. If the gas is unpolarized, i.e., 
$C_{\lambda}=1,\,\forall \lambda$, the energy per particle reduces to the known 
expression~\cite{bishop},
\begin{equation}
\frac{E}{N}  =   \frac{3 \epsilon_F}{5} \left\{ 1 + (\nu -1) \left[ 
\frac{10}{9 \pi} x 
 +   \frac{20}{105 \pi^2} (11 - 2 \ln 2) \, x^2 \right] \right\} \ .
\label{enernopol}
\end{equation}

The energy of the interacting Fermi gas (\ref{ener2}) is written in terms of 
the occupation of the different spin channels $C_{\lambda}$. This set of 
values results in a polarization $P$ of the system, in such a 
way that when all the spin states are equally populated the gas is unpolarized 
$P=0$ and, if only one of them is populated, $|P|=1$. Keeping the total number of 
particles $N$ as constant, and for $s=1/2$, there is only one ratio of spin 
occupations for a given value of $P$. In contrast, for $s>1/2$ there are more 
combinations. It can be shown that the solution which optimizes the energy is the 
one in which the increase of particles in one spin state comes for an equal 
decrease of the rest with constant $N$~(See App. \ref{appendix:a}). Under these conditions, 
the 
concentrations $C_{\lambda}$ for a given polarization $P$ are 
\begin{eqnarray}
C_+ & = & 1+ |P| \, (\nu-1) 
\label{cesp2}\\
C_{\lambda \ne +} & = & 1-|P| \ ,
\label{cesp}
\end{eqnarray}
with subindex $+$ standing for the state with the larger population. As we have discussed, for $s>1/2$ there are more possible configurations. For example, Ref.~\cite{prethermalization} works with a system of $s=3/2$ with population imbalance: the number of atoms with $s=3/2$, $N_{\pm3/2}$, is different than the one with $s=1/2$, $N_{\pm1/2}$. As we need to make a choice, we choose the one that minimizes the energy.

It is interesting to check if Eq. (\ref{ener2}) converges when the degeneracy increases. If we set the limit $\nu\rightarrow\infty$ in Eq. (\ref{ener2}), we get
\begin{equation}
\begin{aligned}
\frac{E}{N}=\frac{2\pi\hbar^2}{ma_0^2}\bigg[\frac{3}{20\pi}(6\pi^2n)^{2/3}a_0^2P^{5/3}+na_0^3(1-P^2)\\
+\frac{1}{2\pi^3}(6\pi^2n)^{4/3}a_0^4(1-P)P^{4/3}\bigg] \ .
\end{aligned}
\label{enevinf}
\end{equation}
As one can see, Eq. (\ref{enevinf}) is convergent, as it is finite for any value of $P$. Particularizing to the unpolarized case $P=0$, one obtains the Hartree-Fock energy for bosons. And, if we set $P=1$, we obtain the energy of a non-interacting Fermi gas.

\section{Landau Theory}
Upon an increase of the gas parameter, the interacting Fermi gas will eventually become
ferromagnetic. To localize the transition point and the order of the phase 
transition we rely on Landau theory. We first explore what we obtain for the Stoner model, that is, first order approximation in $x$. Then, we consider the second-order expansion, Eq. (\ref{ener2}). In this way,  we can compare the differences that arise due to increasing the order of the perturbative expansion. At the Hartree-Fock level, first order in 
$x$, the Landau expansion of the energy is given by
\begin{equation}
    f-f_0(x)=-\frac{1}{2}A(x-x_0)P^2-\frac{1}{3}B{|P|}^3+\frac{1}{4}CP^4 \ ,
    \label{eq.landauStoner}
\end{equation}
with $f= 5E/(3N\epsilon_F)$, $x_0=\pi/2$, and the rest of constants are given 
in App. \ref{appendix:b}. In order to determine the transition point, one 
imposes two 
conditions: \textit{i})  the energy has to be a minimum, that is, its first 
derivative must be zero, and  \textit{ii}) the energy must be smaller than the 
zero-polarization energy to have a global minimum. With these two criteria, one 
finds $x^{\ast}=x_0-2B^2/(9AC)$ and $|P^{\ast}|=2B/(3C)$. The constant $B$ is 
proportional to $(\nu-2)$, therefore the Stoner model predicts a first-order 
phase transition for spin higher than 1/2, while a continuous one for  
$s=1/2$.

Equipped with the second-order expression of the energy as a function of the 
spin-state occupations (\ref{ener2}), we can make a Landau expansion of the 
energy. We get
\begin{equation}
\begin{aligned}
    f-f_0(x)=-\frac{1}{2}A(\overline{x}(x)-x_0) P^2-\frac{B(x)}{3}{|P|}^3
    \\+\frac{C(x)}{4}P^4+\frac{L(x)}{4}P^4\ln{|P|} \ .
\end{aligned}
    \label{Elandauv}
\end{equation}
The coefficients in Eq. (\ref{Elandauv}) depend now on the gas parameter $x$ and 
have a more complex expression (See App. \ref{appendix:b}). For the sake of simplicity, from now on, we will not specify that the coefficients of Eq. (\ref{Elandauv}) depend on the gas parameter $x$. Applying 
the same criteria for finding the transition point as before, 
we derive the equations for the jump of polarization and transition density values,
\begin{equation}
    \ln{|P^{\ast}|}=-\bigg(\frac{1}{2}+\frac{C}{L}\bigg)+\frac{2B}{3L|P^{\ast}|}
    \label{pstar2nd}
\end{equation}
\begin{equation}
   \overline{x}^{\ast}=x_0-\frac{B}{3A}|P^{\ast}|-\frac{L}{4A}|P^{\ast}|^2 \ .
\label{xstar2nd}
   \end{equation}
We rearrange Eq. (\ref{pstar2nd}) to explore the range where
a solution exists,
\begin{equation}
    \frac{2B}{3L|P^{\ast}|}-\ln{|P^{\ast}|}=\frac{1}{2}+\frac{C}{L} \ .
\end{equation}
The function $2B/(3L|P^{\ast}|)-\ln{|P^{\ast}|}$ is continuously decreasing 
between $|P^{\ast}|=0$ and $|P^{\ast}|=1$ as long as $2B/(3L)$ is positive. As 
this is the case, in this range, the minimum value will be at $|P^{\ast}|=1$. 
Hence, there will exist a solution when the coefficients satisfy the condition
\begin{equation}
    \frac{1}{2}+\frac{C}{L}-\frac{2B}{3L}\geq 0 \ .
\end{equation}
This condition is fulfilled for any spin value and for densities 
lower than the transition one. For spin 1/2, the above equations suffer an 
important simplification as the coefficient $B$ is zero,  
\begin{equation}
\begin{aligned}
    |P^{\ast}|=\exp{\bigg(-\frac{1}{2}-\frac{C}{L}\bigg)}\\
    \overline{x}^{\ast}=x_0-\frac{L}{4A}\exp{\bigg(-1-\frac{2C}{L}\bigg)} \ .
\end{aligned}
\label{pstar12}
\end{equation}
It is worth noticing the relevance of the coefficient $L$ in Eq. 
(\ref{pstar12}), coming from the term $P^4 \ln|P|$, for distinguishing the 
order of the phase transition: $L$ needs to be different from zero to get a 
first-order phase transition~\cite{lianyi}.
Introducing the values of the constants in Eq. (\ref{pstar12}), one gets 
 the 
transition density at $x=\pi/3(1+0.007)\approx1.0545$, and a jump in the 
polarization equal to $0.545$. Remarkably, the second-order energy changes the 
character of the ferromagnetic phase transition becoming now a first-order one, 
as it happens for spin $s>1/2$.

\section{Results}
We have applied the above formalism to study fundamental properties of the
Fermi gas, such as the energy or the magnetic susceptibility.

\begin{figure}[h]
    \centering
    \includegraphics[width=0.9\linewidth]{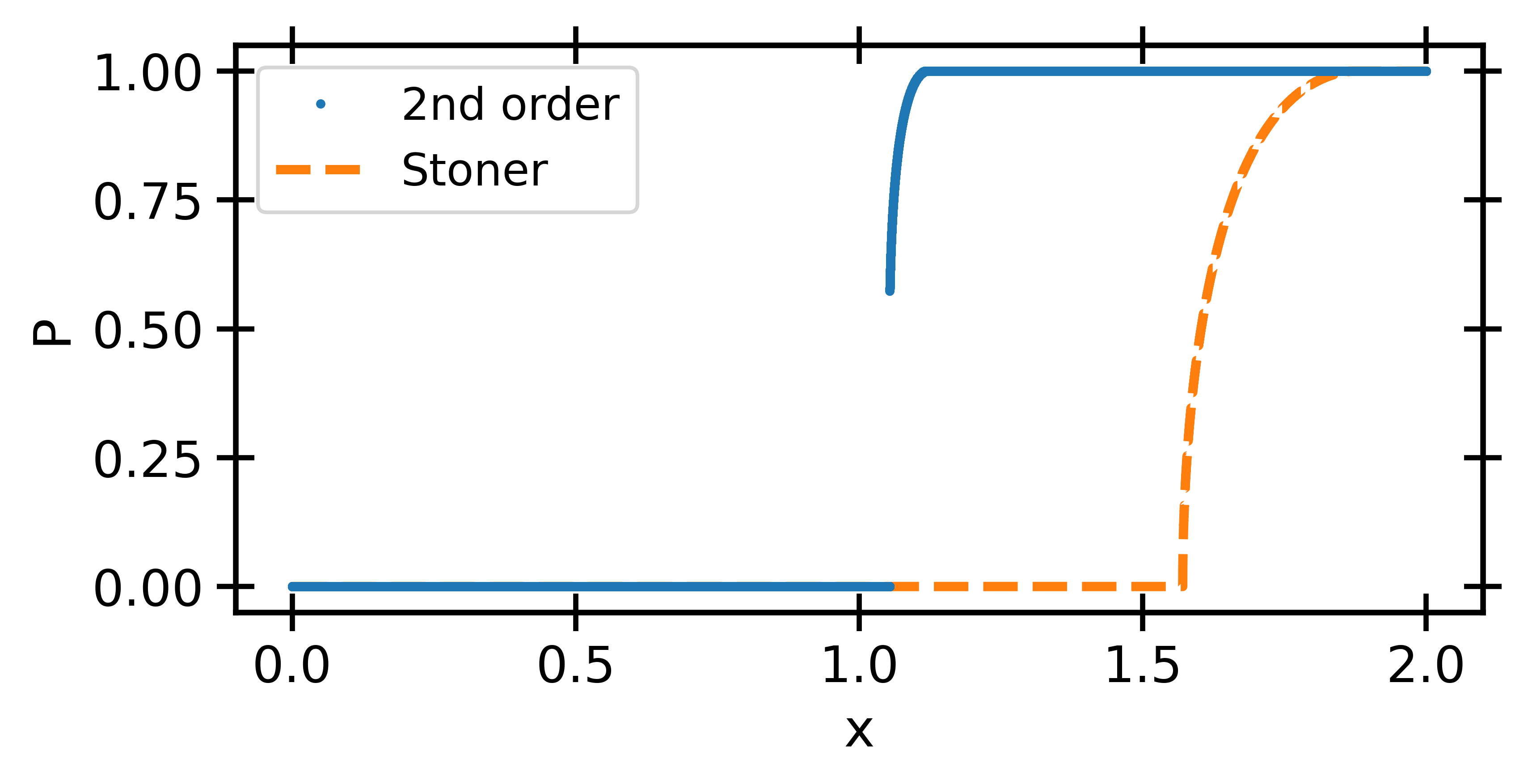}
    \caption{Itinerant ferromagnetic transition for $s=1/2$. The  
dashed orange and solid blue lines stand for the Stoner and second-order approximations, 
respectively.}
    \label{trans1-2}
\end{figure}

In Fig.\ref{trans1-2}, we show the dependence of the polarization $P$ with the 
gas parameter $x$ for a $s=1/2$ Fermi gas. One can observe as the introduction 
of second-order corrections to the potential energy modify the character of the 
phase transition with respect to the Stoner model. Moreover, the gas parameter 
at which the transition occurs  is significantly reduced, from $\pi/2$ to 
$\sim \pi/3$, approaching the value $x \simeq 1$ where the transition has been indirectly observed~\cite{valtolina}.  

\begin{figure}[h]
    \centering
    \includegraphics[width=0.9\linewidth]{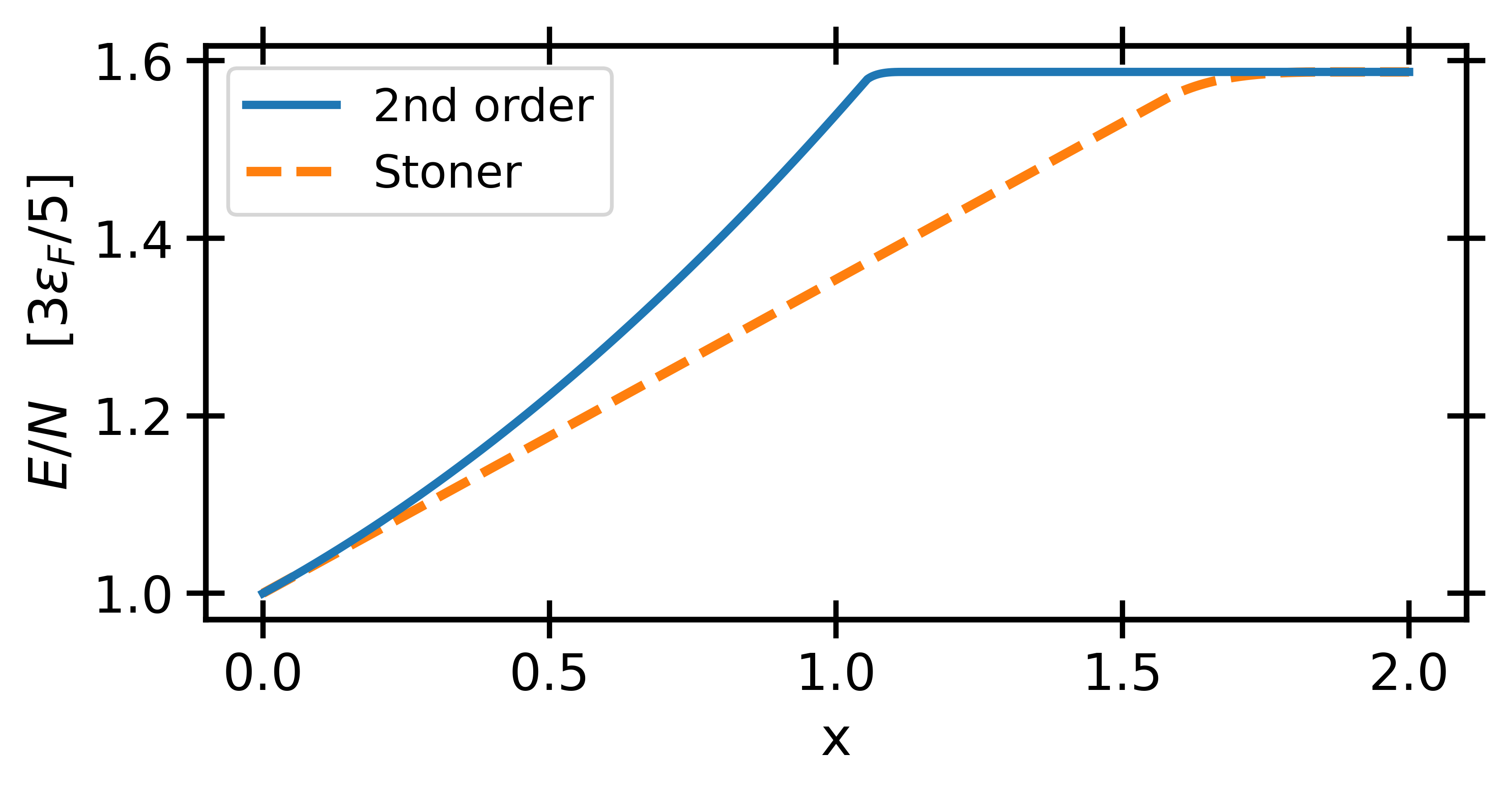}
    \caption{Energy per particle as a function of the gas parameter for $s=1/2$. The dashed orange and solid blue lines stand for the Stoner and second-order approximations, respectively.}
    \label{fig:edv2}
\end{figure}

In Fig. \ref{fig:edv2}, we plot the energy of the Fermi gas using the 
the Stoner one and the second-order approximation. At very low 
densities, both models predict the same energy, which follow the non-polarized 
behavior. After a certain value of the gas parameter, the lines become flat, 
which is the behavior of the fully-polarized gas. We can see that, when we use 
the second-order approximation, the transition happens at a smaller gas 
parameter.

\begin{figure}[tb]
    \centering
    \includegraphics[width=0.9\linewidth]{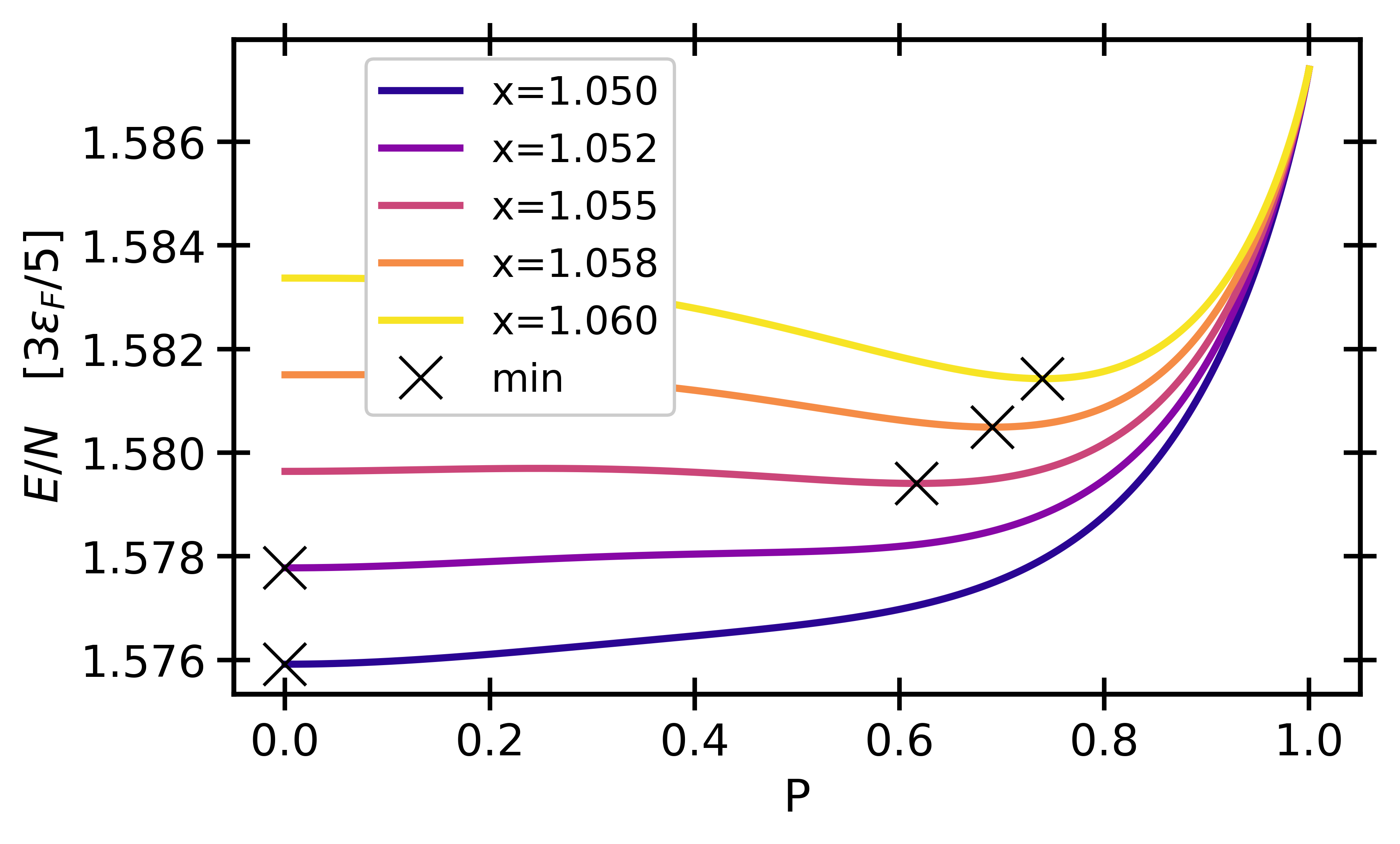}
    \caption{Energy per particle as a function of the polarization $P$ for 
$s=1/2$. The lines correspond to different $x$ values close to the 
phase transition $x^\star =1.054$. The crosses in each line indicates the polarization where the energy is minimum.}
    \label{energy1-2}
\end{figure}

The first-order phase transition for $s=1/2$ is also observed in Fig. 
\ref{energy1-2}. There, we plot the energy as a function of $P$ and for gas 
parameter values close to the phase transition. One can see that the minimum of 
the energy jumps from $P=0$ to an intermediate value $P=0.545$ that then 
progressively moves to the fully polarized phase (Fig. \ref{trans1-2}).

\begin{figure}[h]
    \centering
    \includegraphics[width=0.9\linewidth]{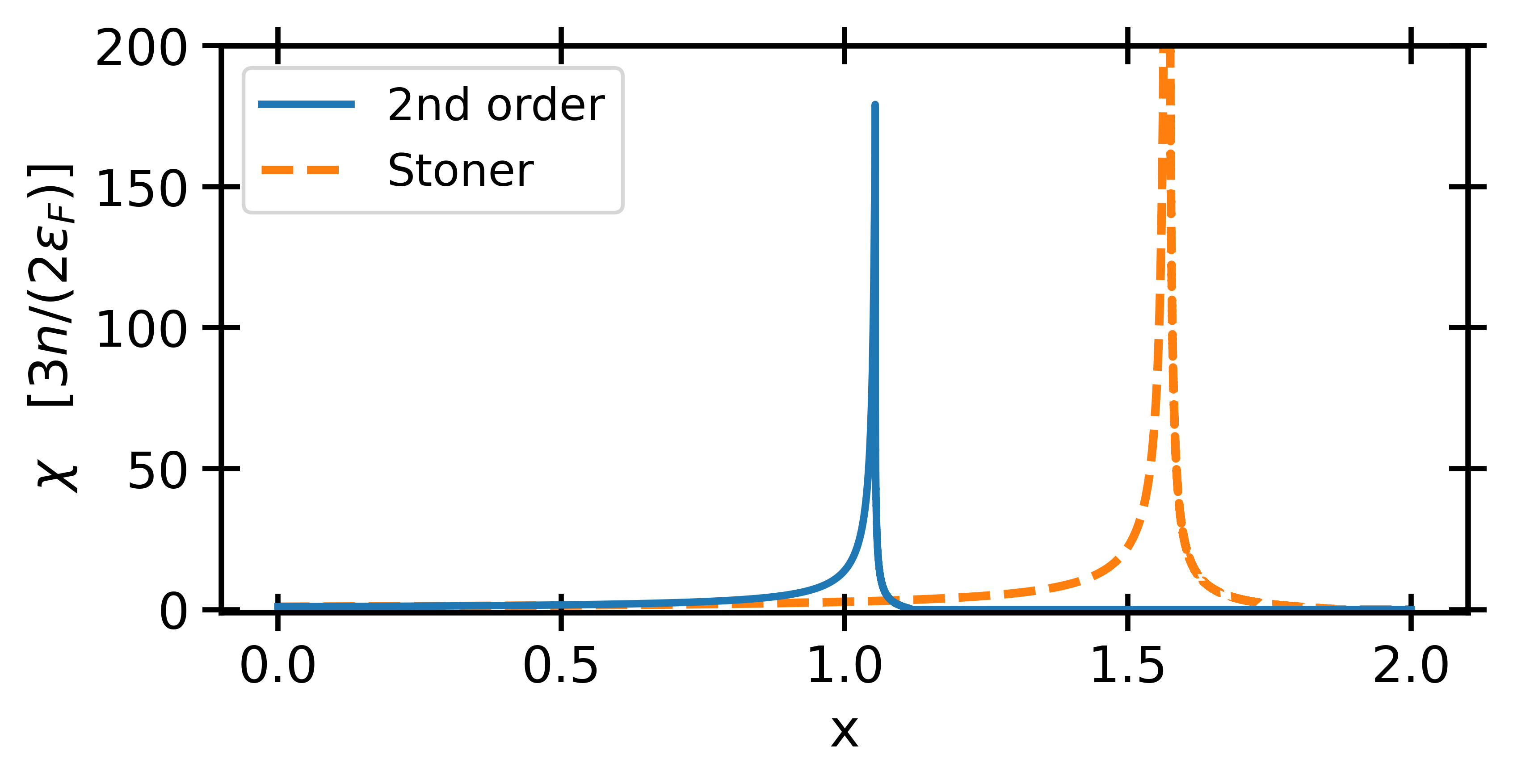}
    \caption{Magnetic susceptibility $\chi$ around the transition point for $s=1/2$. The peak of $\chi$ appears at a lower $x$ value than in the Stoner model.}
    \label{chi1-2}
\end{figure}

Since we are interested in the magnetism of these gases, we proceed to analyze the magnetic 
susceptibility, which is inversely proportional to the second derivative of the 
energy with respect to the polarization,
\begin{equation}
   \frac{1}{\chi}=\frac{1}{n}\left(\frac{\partial^2 (E/N)}{\partial 
P^2} \right)_x \ .
\label{suscept}
\end{equation}
If we split the total energy between the kinetic and the potential energy, we 
can rewrite $\chi$ as
\begin{equation}   
\chi=\frac{3}{2}\frac{n}{\epsilon_F}  \left[ \frac{1}{\nu}\sum_{\lambda}C_{
\lambda}^{-1/3}(C_{\lambda}')^{2}+\frac{3}{2}\frac{1}{\epsilon_F} 
(V/N)'' \right]^{-1}  \ ,
\label{suscept2}
\end{equation}
where the derivatives are with respect to the polarization. In Fig. 
\ref{chi1-2}, we show the magnetic susceptibility around the ferromagnetic 
transition point for $s=1/2$. We see again that the transition occurs at a lower 
value of $x$ \cite{pilati} with respect to the Stoner model. One can also 
notice that $\chi$ 
changes behavior, from diverging at the transition point (Stoner) to a large 
but finite peak 
at second order, reflecting the change in the type of phase transition from a continuous 
to a first-order one. 

From Eq. (\ref{suscept2}), one can prove that at first order $\chi$ diverges. At this order,  $\chi$ behaves around the transition as
\begin{equation}
    \chi=\frac{3}{2}\frac{n}{\epsilon_F}  \left[-\frac{9}{10}A(x-x_0) \right]^{-1}=\frac{3}{2}\frac{n}{\epsilon_F}\frac{5C}{B^2} \ .
    \label{suscept3}
\end{equation}
The coefficient $B$ in Eq. (\ref{suscept3}) is proportional to $(\nu-2)$ and $C$ is finite for $\nu=2$ (See App. \ref{appendix:b}), hence, the magnetic susceptibility diverges around the transition for the Stoner model.

\begin{figure}[h]
    \centering
    \includegraphics[width=0.9\linewidth]{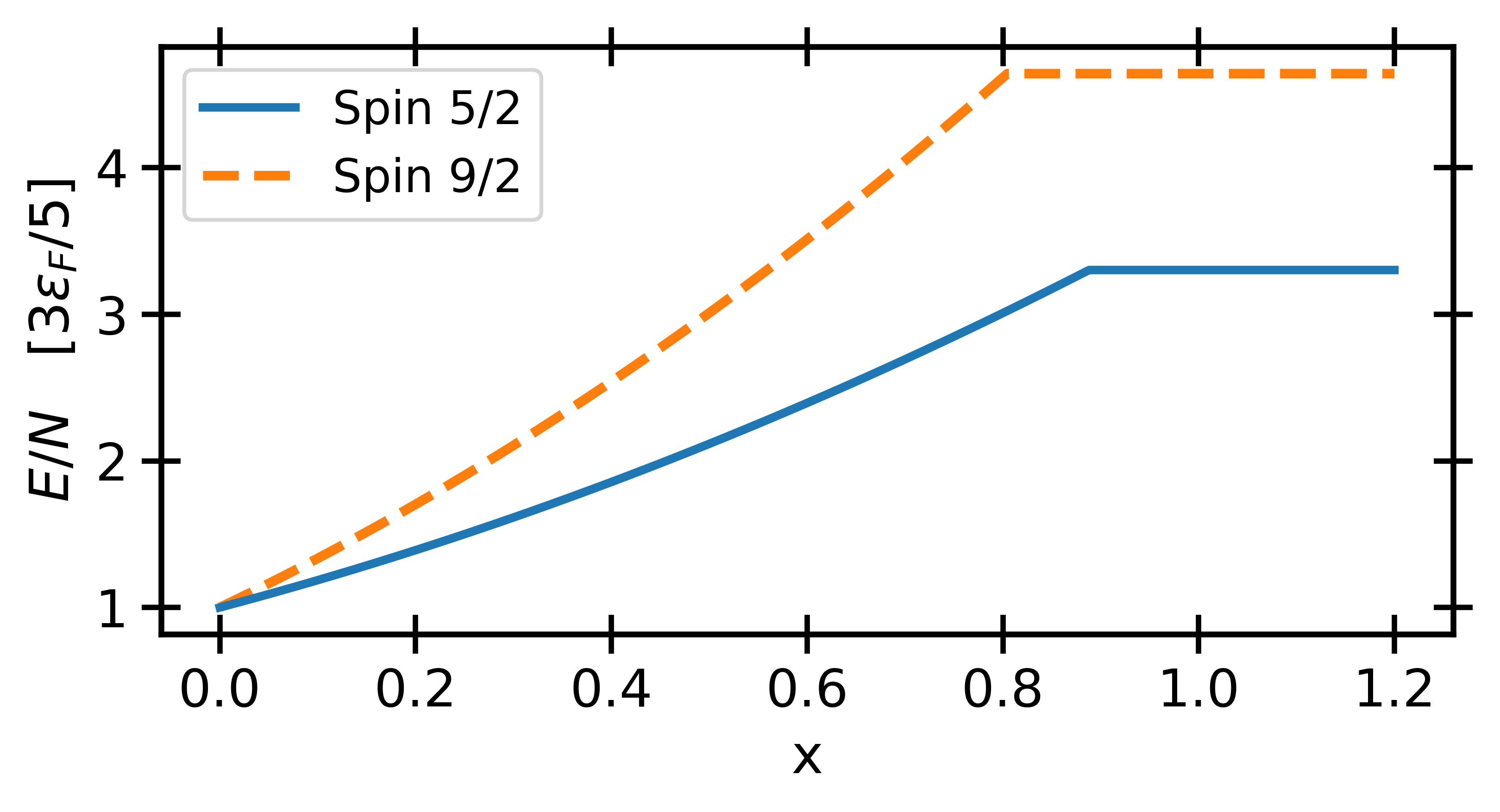}
    \caption{Energy per particle as a function of the gas parameter. The solid blue and dashed orange lines stand for a Fermi gas of $s=5/2$ and $s=9/2$, respectively. Both lines have been calculated within the second-order approximation.}
    \label{fig:edv6v10}
\end{figure}

We can extend our analysis to larger spin Fermi gases. In this case, the 
ferromagnetic transition is first-order as in the Stoner model but the 
transition point is, in all cases, observed at smaller $x$ values. This fact can be seen in Fig. \ref{fig:edv6v10} where the energy of the spin 9/2 Fermi gas changes its behavior at a lower gas parameter than the energy of the spin 5/2 Fermi gas.

\begin{figure}[tb]
    \centering
    \includegraphics[width=0.9\linewidth]{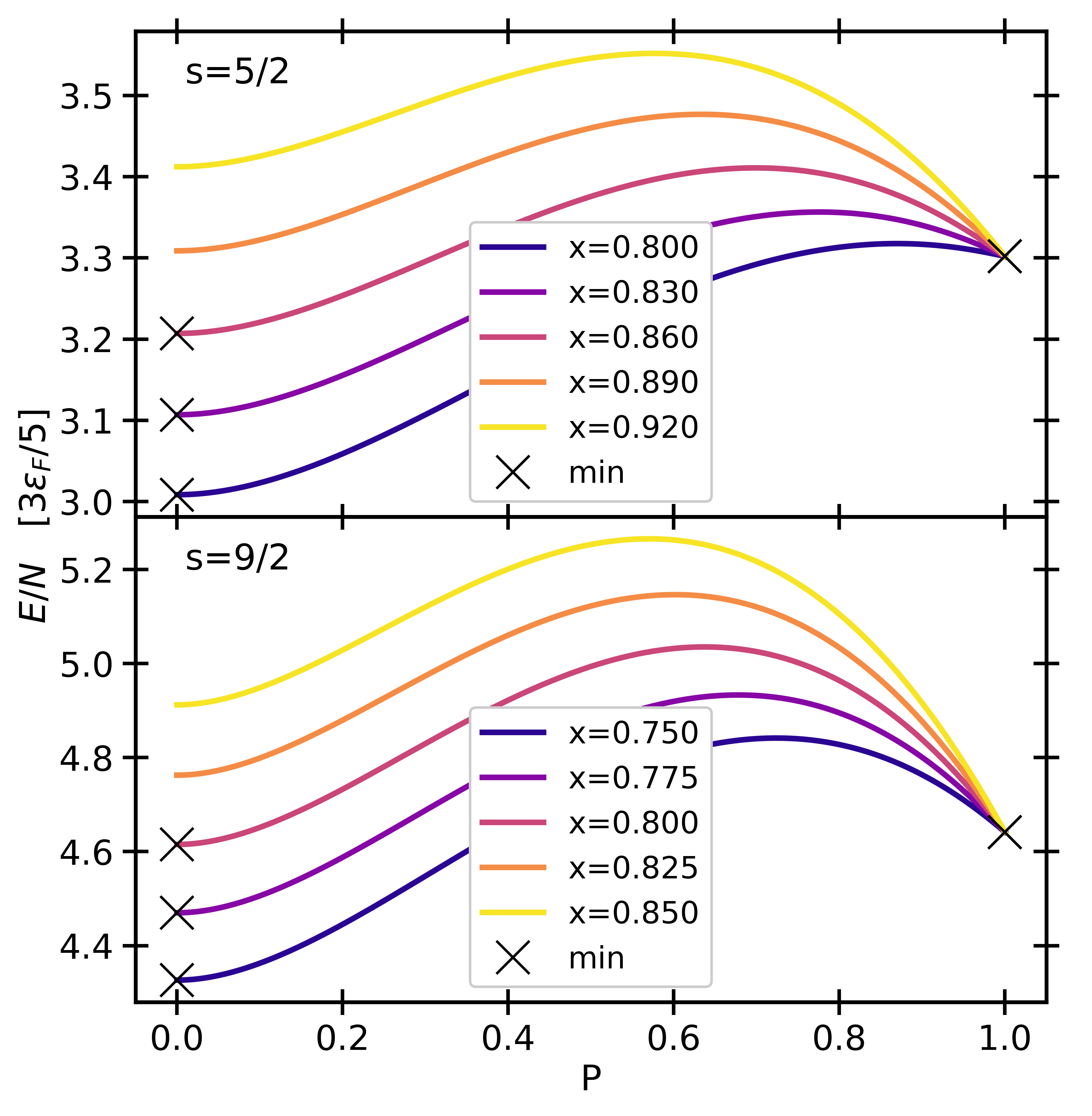}
    \caption{Energy of the $s=5/2$ (top) and $s=9/2$ (bottom) Fermi gases as a 
function of $P$ for $x$ values close to the transition. The crosses indicate the values of $P$ where the energy is minimum.}
    \label{energysun}
\end{figure}

\begin{figure}[h]
    \centering
    \includegraphics[width=0.9\linewidth]{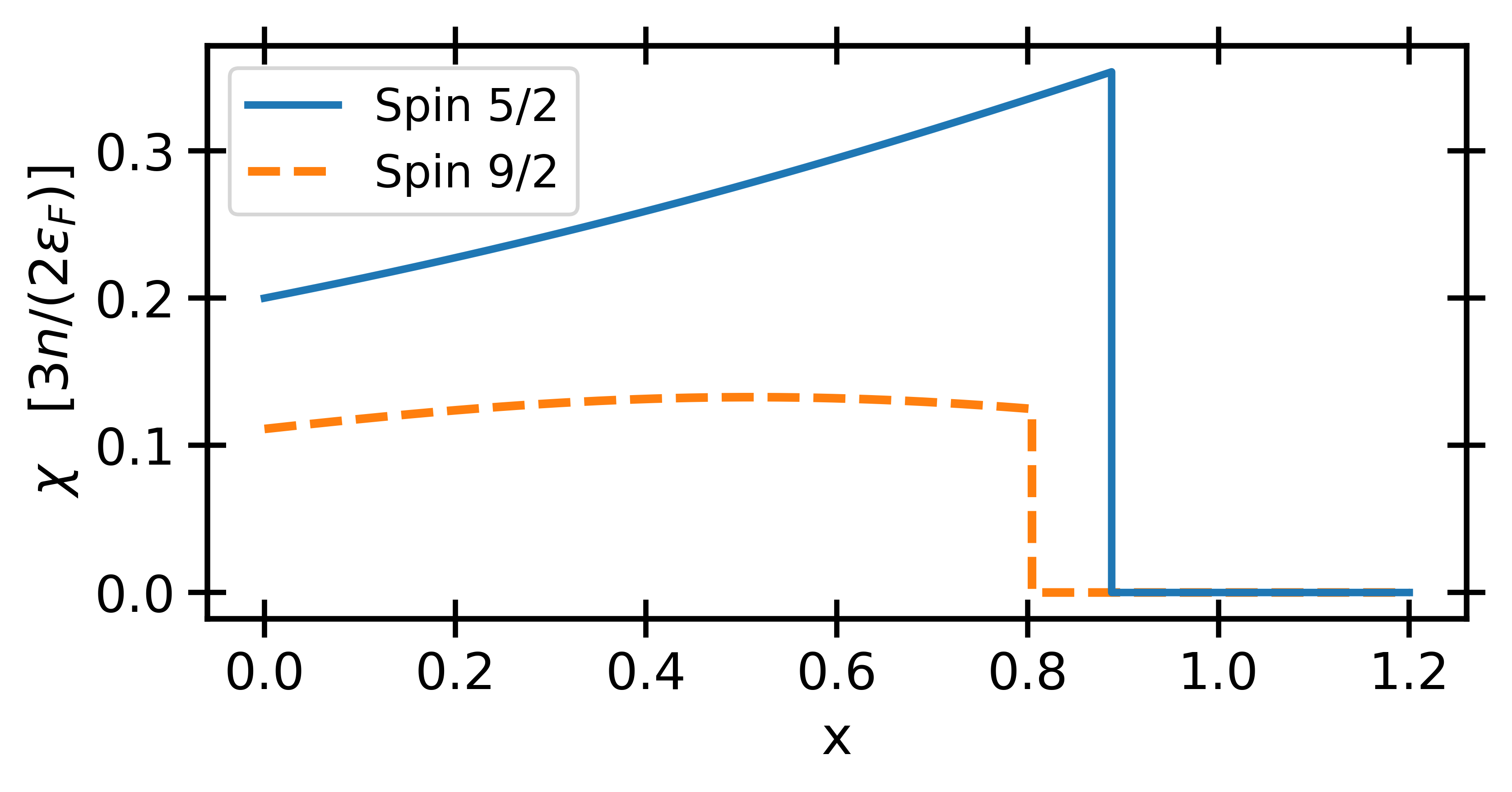}
    \caption{Magnetic susceptibility per particle as a function of the gas parameter. The slid blue and dashed orange lines stand for a Fermi gas of $s=5/2$ and $s=9/2$, respectively. Both lines have been computed with the second-order approximation.}
    \label{fig:chidv6v10}
\end{figure}

 In 
Fig.~\ref{energysun}, we show the energies of SU(N) Fermi gases up to second 
order as a function of the polarization for $x$ values close to the phase 
transition. The top panel is for  $s=5/2$ and the bottom one for $s=9/2$ 
corresponding to Yb and Sr, respectively. In both cases, the location of the minimum of the energy as a function of the polarization jumps 
from $P=0$ to 1 without intermediate values, in contrast to the case of 
$s=1/2$. For the sake of completeness, in Fig. \ref{fig:chidv6v10} we show the 
magnetic susceptibility for the two high degenerate Fermi gases considered: 
$s=5/2$ and $s=9/2$.

Thanks to having an analytical expression for the energy, we have access to 
other important properties as the Tan's 
contact. According to Tan relations, the microscopic behavior of the wave 
function of the $N$-body system at short distances ($r \ll n^{-1/3}$) is 
connected with 
the behavior of several macroscopic magnitudes. In particular, the tail of the 
momentum distribution for large $\textbf{k}$ values as $k^{-4}$ and the 
dependence of the energy on the scattering length. Using the adiabatic sweep theorem~\cite{abiatictheorem}, the Tan's 
contact can be obtained from the energy of the system by
\begin{equation}
   C=\frac{8\pi ma_0^2}{\nu\hbar^2}\frac{N}{V}\frac{\partial 
(E/N)}{\partial a_0} \ .
\end{equation}
In terms of the gas parameter,
\begin{equation}
C=\frac{4\pi nk_F}{\nu}\frac{x^2}{\epsilon_F}\frac{\partial (E/N)}{\partial x} \ .
\end{equation}

\begin{figure}[h]
    \centering
    \includegraphics[height=0.52\linewidth,width=0.9\linewidth]{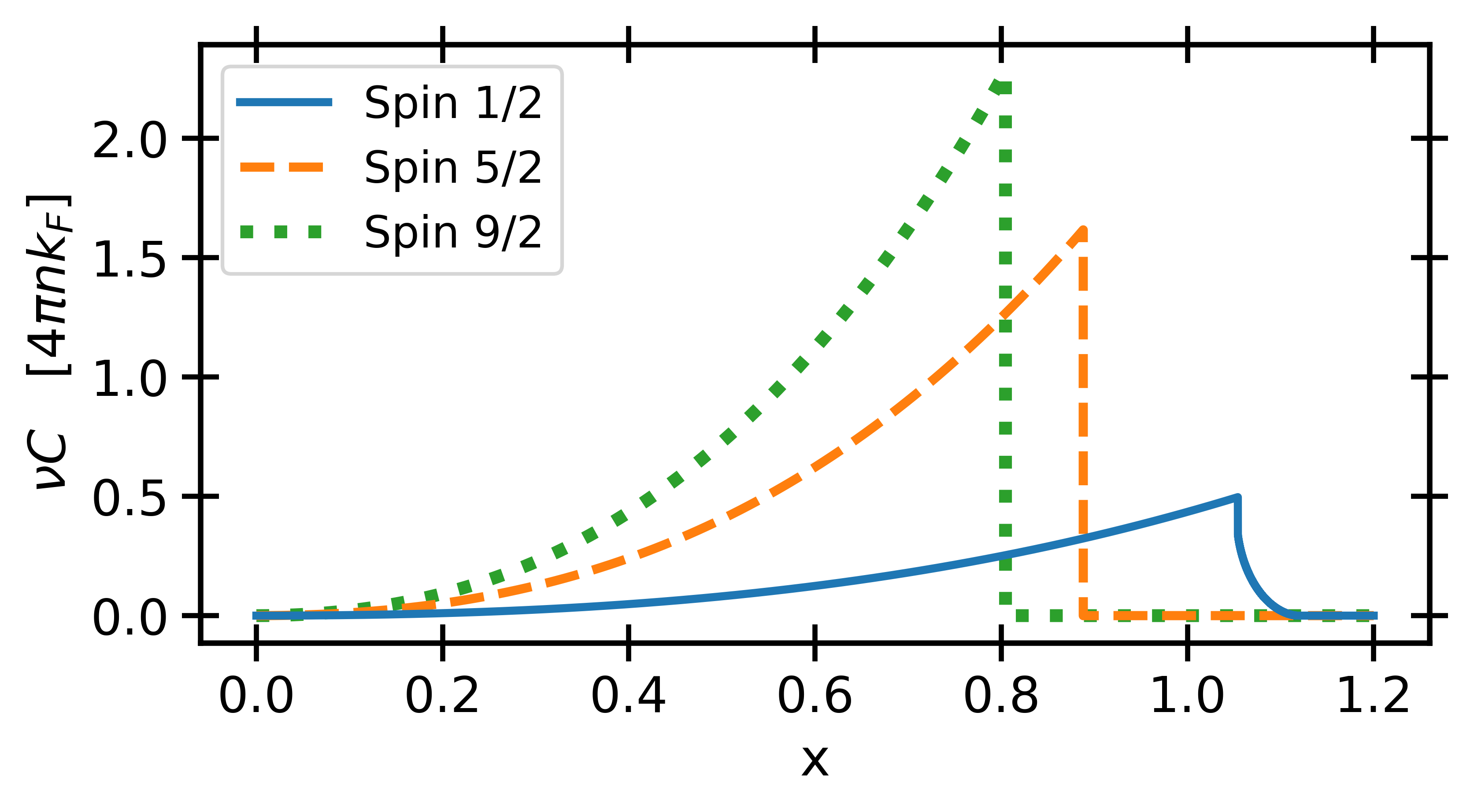}
    \caption{Tan's contact in SU(N) Fermi gases. $\nu C$ as a 
function of the gas parameter for three different spins: $s=1/2$, $5/2$, and $9/2$. We use the second-order model for the three cases.}
    \label{tanfig1}
\end{figure}

We have calculated the Tan's contact using the energy of the Fermi gas at second 
order. In Fig. \ref{tanfig1}, we show  $\nu C$ as a function of $x$ for spin 
$s=1/2$, $5/2$, 
and $9/2$. In the paramagnetic phase, $\nu C$ increases monotonically until it 
reaches its maximum value at the transition point. After crossing the ferromagnetic 
transition, the Tan's contact becomes zero because the energy of the fully 
polarized phase does not depend on the scattering length $a_0$. As one can see 
in the figure, the $\nu C$ increases with the value of the spin for a given
$x$ value, in agreement with the increase of interaction energy with spin degeneracy. 
It is interesting to notice that, for spin $1/2$, $C$ does not drop abruptly 
to zero because there are stable polarizations between 0 and 1. In contrast, for 
$s>1/2$ the drop is directly to zero because of the sudden change of the polarization
from 0 to 1 at the phase transition.

\begin{figure}[h]
    \centering
    \includegraphics[width=0.9\linewidth]{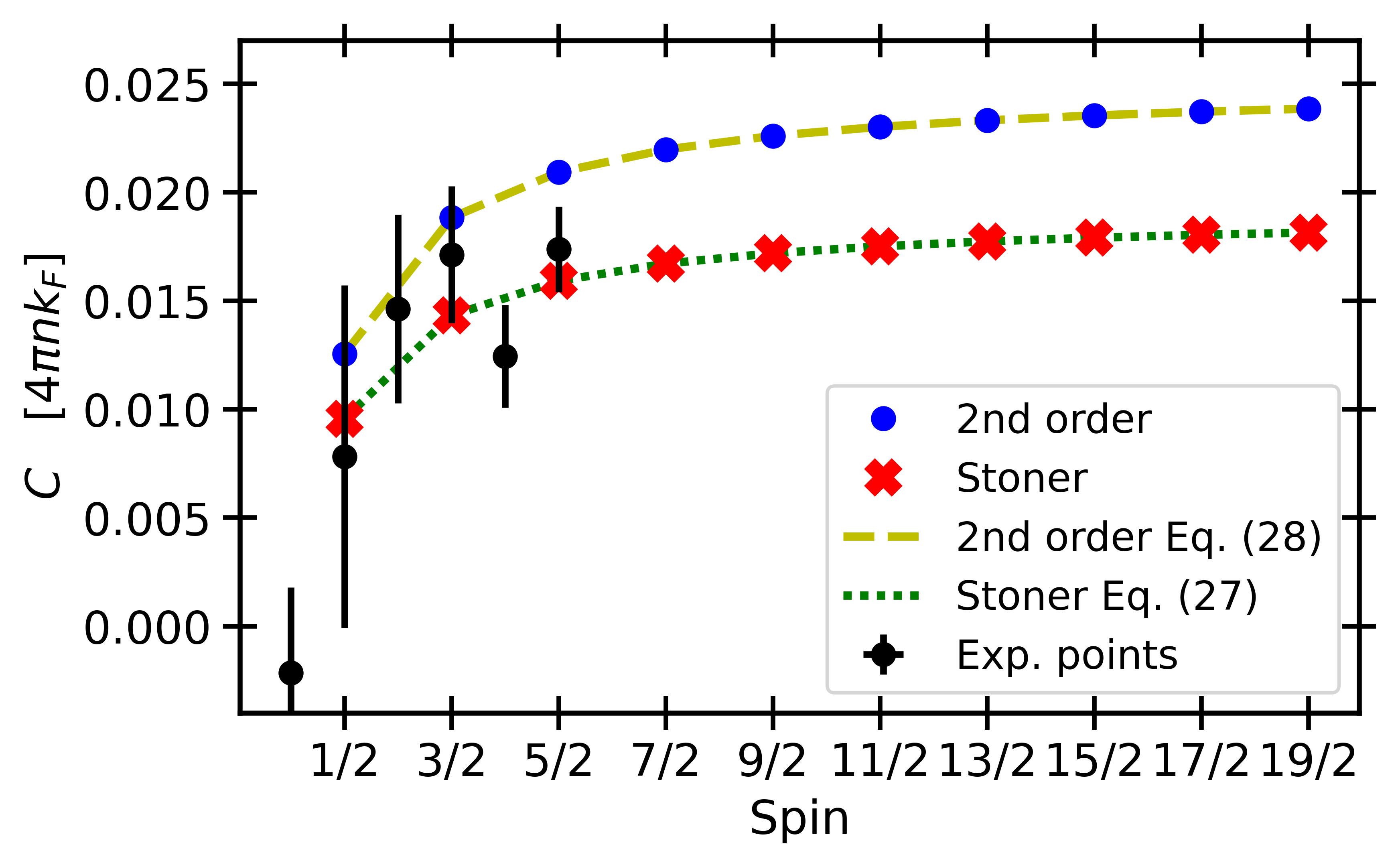}
    \caption{Tan's contact in SU(N) Fermi gases. Evolution of $C$ with the spin of particles at fixed density $x=0.3$ for the Stoner model and the second order model; the points 
with error bars are experimental data from Ref.~\cite{song} re-scaled by a factor of $(2\pi)^2$. The green and yellow lines follow the behavior of the Tan's contact of Eq. (\ref{contact1}) and Eq. (\ref{contact2}) respectively. }
    \label{tanfig2}
\end{figure}

In Fig.~\ref{tanfig2}, we show the evolution of the Tan's contact as a function of the spin for the Stoner and second-order models. By increasing the spin, one can see a 
tendency to reach a plateau. This 
plateau, which has been interpreted as the Bose limit, can be understood by looking at the Tan's contact behavior before the transition. For the Stoner model, it is given by
\begin{equation}
    C=4\pi n k_F x^2\frac{2}{3\pi}\bigg(1-\frac{1}{\nu}\bigg) \ ,
    \label{contact1}
\end{equation}
and in second-order,
\begin{equation}
    C=4\pi n k_F x^2\bigg[\frac{2}{3\pi}+\frac{8}{35\pi^2}(11-2\ln{2})x\bigg]\bigg(1-\frac{1}{\nu}\bigg) \ .
    \label{contact2}
\end{equation}
The dependence with $\nu$ for both models is $1-1/\nu$, hence, if we set the limit $\nu\rightarrow\infty$, we obtain a plateau.
In Fig.~\ref{tanfig2}, we also plot 
experimental results from Ref. \cite{song} that show a similar behavior to our theoretical expressions (Eqs. (\ref{contact1}) and (\ref{contact2})). Moreover, they lie between both models, but we cannot state which one fits better due to the experimental uncertainty. As the experimental points have a different normalization than our definition, we have to scale them by a factor of $(2\pi)^2$.

\begin{figure}[tb]
    \centering
    \includegraphics[width=0.9\linewidth]{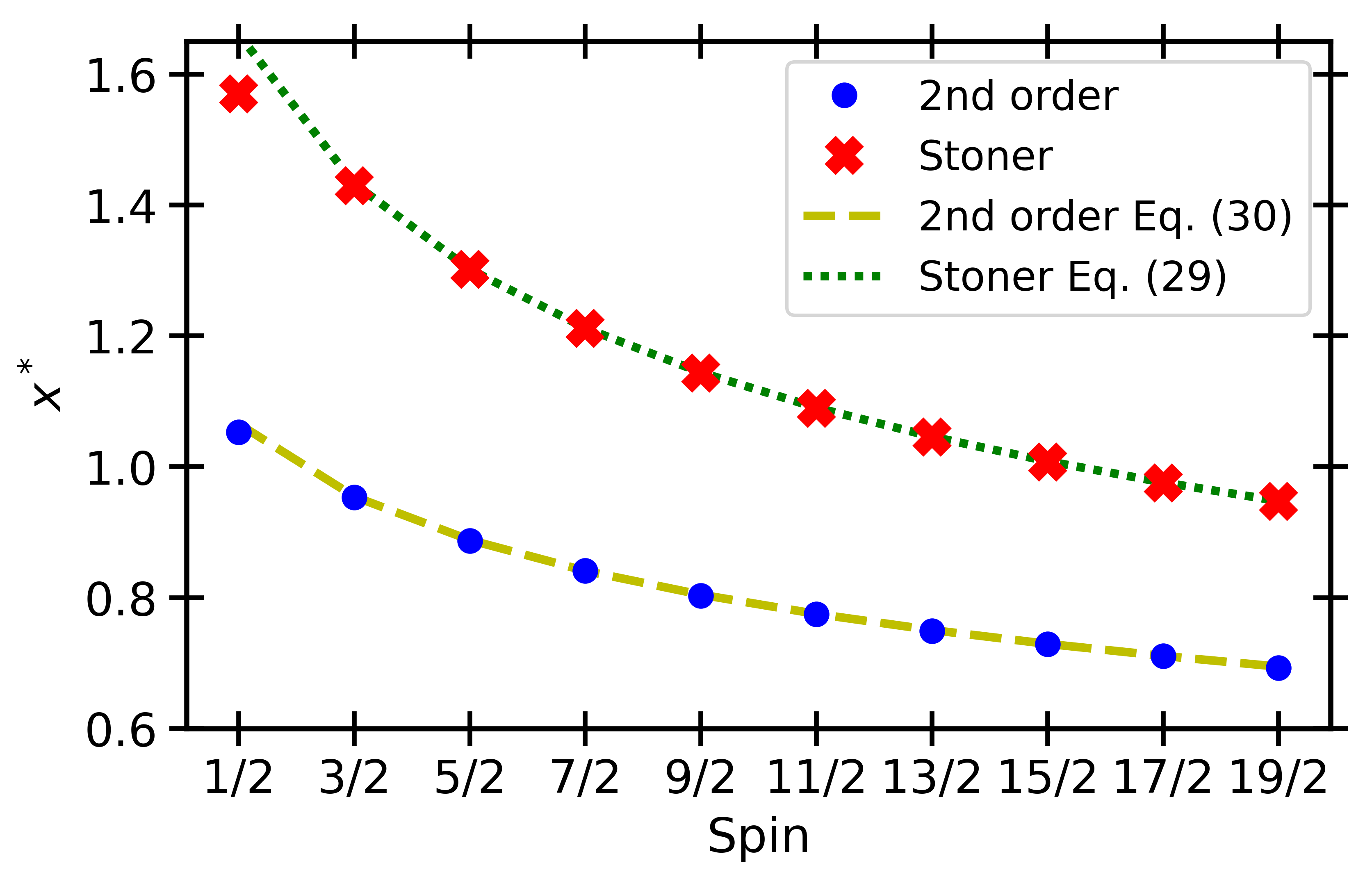}
    \caption{Critical gas parameter values of the ferromagnetic transition as a function of the spin for the Stoner model and the second order model. The green and yellow lines follow the behavior predicted by Eqs. (\ref{critic1}) and (\ref{critic2}).}
    \label{transisun}
\end{figure}

The critical value of the gas parameter depends on the spin of the particles. 
In Fig.~\ref{transisun}, we report the results obtained for both models (Stoner and second order). As one 
can see, the itinerant ferromagnetic transition happens at $x$ values that 
decrease monotonically with the spin degeneracy. Moreover, the second order values of $x$ are lower than the ones predicted by the Stoner model. As for $s>1/2$ the polarization goes from 0 to 1 without intermediate values ($P^*=1$), we can find more easily the laws that the critical values $x^*$ follow. For the Stoner model, the law is given by
\begin{equation}
    x^*=\frac{9\pi}{10(\nu-1)}\big(\nu^{2/3}-1\big) \ ,
    \label{critic1}
\end{equation}
and in second-order,
\begin{equation}
    x^*=\pi\frac{-35+35\sqrt{1+\cfrac{108(11-2\ln{2})}{175(\nu-1)}\big(\nu^{2/3}-1\big)}}{12(11-2\ln{2})} \ .
    \label{critic2}
\end{equation}
Notice that for $s=1/2$ Eqs. (\ref{critic1}) and (\ref{critic2}) do not hold. In this case, 
the value of $x^*$ is slightly smaller than the value predicted by the
laws (\ref{critic1},\ref{critic2}) because the transition is from $P=0$ to  $P<1$ (see Fig. \ref{transisun}).
The behavior of the critical gas parameter values with the spin, in Fig. \ref{transisun}, can be 
understood by the balance between Fermi kinetic energy and interaction energy. 
Increasing the spin degeneracy originates in turn an increase in the number of 
interacting pairs of particles
because the interatomic potential acts only between pairs of different $z$-spin 
component. In other words, the Fermi gas becomes more interacting and reaches 
the fully polarized (ferromagnetic) phase at lower $x$ values.

\section{Conclusions}
 
Summarizing, we present the analytic expression of the energy of a repulsive 
SU(N) Fermi gas, in terms of the spin-channel occupations, at second order of 
the 
gas parameter. This analytic derivation allows for an accurate estimation of 
the magnetic properties of the Fermi gas for any value of the spin. Moreover, 
 by using the analytical solution one directly avoids any uncertainty 
coming from the numerical integration. This is in fact  quite important since 
the function to integrate in the second-order term has many singular points and, 
if one does not apply a previous analytical treatment to the function, the 
numerical method may simply diverge. In order to study the system, we have chosen the occupational configuration and the polarization protocol that minimizes the energy: at $P=0$, all the species are equally occupied; and, when we increase $P$, one species increases and all the rest diminish in the same manner. However, we point out that our formalism can be applied to any occupational configuration and any polarization protocol. One just needs to find the new expressions for the fractions of $\lambda$ particles $C_{\lambda}$. In fact, in many experiments, the species' occupation can be tuned or controlled almost at will~\cite{collective_excitations,estronci_N_10} and there are studies that have dealt with imbalanced systems \cite{prethermalization}. However, we point out that all these other configurations are excited states, they have more energy than the configuration we have chosen. Due to that, if a Fermi system can thermalise, it will fall to the behavior we have predicted in this work. Using another occupational configuration and a different polarization protocol leads to different critical values of the gas parameter $x^*$.

Our results show that the ferromagnetic transition turns out to be first-order for $s=1/2$, in 
contrast with the continuous transition obtained at the Hartree-Fock (Stoner) 
approximation~\cite{stoner}. At second order, and for $s>1/2$, the phase 
transition is always of first-order type. 
Our derivation applies to any spin of the Fermi 
particles and 
this allows for the study of itinerant ferromagnetism in SU(N) fermions. The 
critical gas parameter for spin 1/2 decreases significantly with respect to the 
Stoner model, approaching the experimental estimation $x \simeq 
1$~\cite{valtolina}. Remarkably, 
the critical $x$ decreases monotonically with the spin of the particles 
suggesting that the observation of itinerant ferromagnetism could be favored by 
working with highly-degenerate gases as Yb~\cite{pagano} and Sr~\cite{goban}. In order to verify experimentally our predictions, one could use a similar method to the one used in Ref.~\cite{valtolina} where the Fermi system can thermalise, but generalized to larger spin  degeneracies.

Beyond the second-order 
terms analyzed in this work, the energy ceases to be universal in terms of the 
gas parameters because other scattering parameters of the interactions, mainly 
the $s$-wave effective range and $p$-wave scattering length get involved~\cite{bishop}.
However, these corrections are expected to be small due to the diluteness of the Fermi gases in experiments that make scattering be dominated by $s$-wave scattering.

\section{Acknowledgments}
We acknowledge financial support from   MCIN/AEI/10.13039/501100011033 
(Spain) grant No.  PID2020-113565GB-C21 and from Secretaria 
d'Universitats i Recerca del Departament
d'Empresa i Coneixement de la Generalitat de Catalunya, co-funded by the
European Union Regional Development Fund within the ERDF Operational Program of
Catalunya (project QuantumCat, ref. 001-P-001644).

\appendix
\clearpage
\begin{widetext}
\section{Polarization in SU(N) Fermi gases}
\label{appendix:a}

We have considered that when the occupation of one of the spin channels grows, 
the rest decreases in equal form, keeping the total number of particles as 
constant. This 
results  from the fact that the Hamiltonian does not depend on the angular 
momentum of each species, hence, all the species are equivalent. 
Moreover, this fact can be proved by minimizing the energy,
\begin{equation}
E=\frac{3}{5}N\epsilon_F\frac{1}{\nu}\sum_{\lambda}C_{\lambda}^{5/3}+V(C_{s},...
,C_{-s}) \ ,
\end{equation}
with $\sum_{\lambda}C_{\lambda}=\nu$. 
 We define the channel that grows,  
\begin{equation}
    C_+=\nu -\sum_{\lambda\neq +}C_{\lambda} \ ,
\label{constraint}
    \end{equation}
and  substitute it in the energy,
\begin{equation}
\begin{aligned}
    E=\frac{3}{5}N\epsilon_F\frac{1}{\nu}\bigg[\big(\nu-\sum_{\lambda\neq 
+}C_{\lambda}\big)^{5/3}+\sum_{\lambda\neq 
+}C_{\lambda}^{5/3}\bigg]+V(\nu-\sum_{\lambda\neq 
+}C_{\lambda},C_{s-1}...,C_{-s}) \ .
    \end{aligned}
\end{equation}
Now, we minimize the energy for all the channels, except for the one that we 
have isolated which, due to the constraint (\ref{constraint}), depends on the 
others,
\begin{equation}
    \frac{\partial E}{\partial C_{\sigma}}=0\quad\forall\sigma\neq + \ .
\end{equation}
Then, 
\begin{equation}
\begin{aligned}
\frac{\partial E}{\partial C_{\sigma}}=N\epsilon_F\frac{1}{\nu}\bigg[C_{\sigma}^{2/3}-\big(\nu-\sum_{
\lambda\neq +}C_{\lambda}\big)^{2/3}\bigg]+\frac{\partial V}{\partial 
C_{\sigma}}-\frac{\partial V}{\partial C_{+}}=0 \ .
\end{aligned}
\end{equation}
Rearranging terms, we obtain 
\begin{equation}
    N\epsilon_F\frac{1}{\nu}C_{\sigma}^{2/3}+\frac{\partial V}{\partial 
C_{\sigma}}=N\epsilon_F\frac{1}{\nu}\big(\nu-\sum_{\lambda\neq 
+}C_{\lambda}\big)^{2/3}+\frac{\partial V}{\partial C_{+}} \ .
    \label{eq.minimizeenergy}
\end{equation}
From Eq. (\ref{eq.minimizeenergy}), one can see that all the decreasing 
$C_{\sigma}$ satisfy the same equation, therefore, they must be equal.

\section{Parameters of the Landau expansion}

\label{appendix:b}

In this section we show the expression for all the parameters of the Landau 
expansion in terms of the gas degeneracy ($\nu=2s+1$) and the gas parameter 
($x=k_Fa_0$). We first show the Landau expansion up to fourth order in $P$ 
including the logarithmic term that comes from the second order term in 
perturbation theory.
\begin{equation}
\begin{aligned}    
f-f_0=-\frac{A}{2}(\overline{x}-x_0)P^2-\frac{B}{3}{|P|}^3+\frac{C}{4}P^4+\frac{
L}{4}P^4\ln{|P|} \ .
\end{aligned}
\end{equation}
The parameters that are invariant in any order are
\begin{equation}
\begin{aligned}
    f=\frac{5E}{3N\epsilon_F}\quad;\quad A=\frac{20}{9\pi}(\nu-1)\quad;\quad 
x_0=\pi/2 \ .
\end{aligned}
\end{equation}
We split this section into two subsections. The first one contains the 
expression of the parameters up to first order in perturbation theory 
(Hartree-Fock), and the second one up to second order.

\subsection{1st order parameters}
\begin{equation}
\begin{aligned}
    f_0=1+\frac{10}{9\pi}(\nu-1)x\quad;\quad
    \overline{x}=x\quad;\quad
    B=\frac{5}{27}(\nu-1)(\nu-2)\quad;\quad 
    C=\frac{20}{243}(\nu-1)(\nu^2-3\nu+3)\quad;\quad
    L=0
\end{aligned}
\end{equation}

\subsection{2nd order parameters}
\begin{equation}
\begin{aligned} 
f_0=1+\frac{10}{9\pi}(\nu-1)x+\frac{4}{21\pi^2}(\nu-1)(11-2\ln{2})x^2\quad;\quad
    \overline{x}=x+\frac{2}{15\pi}(22-7\nu+4(\nu-1)\ln{2})x^2\\    
B=\frac{5}{27}(\nu-1)(\nu-2)+\frac{2}{81\pi^2}(\nu-1)(\nu-2)(44+\nu+8(\nu-1)\ln{
2})x^2\\    
C=\frac{20}{243}(\nu-1)(\nu^2-3\nu+3)+\\+(\nu-1)\frac{528-336\nu-16\nu^2+29\nu^3
   -(96-192\nu+128\nu^2-32\nu^3)\ln{2}}{729\pi^2}x^2+\frac{20\nu^3}{243\pi^2}
(\nu-1)\ln{\frac{\nu}{6}}x^2\\
    L=\frac{20\nu^3}{243\pi^2}(\nu-1)x^2\\
\end{aligned}
\end{equation}

For spin 1/2, these expressions reduce to
\begin{equation}
\begin{aligned}
    f_0=1+\frac{10}{9\pi}x+\frac{4(\nu-1)(11-2\ln{2})}{21\pi^2}x^2\quad;\quad
    \overline{x}=x+\frac{8(2+\ln{2})}{15\pi}x^2\quad;\quad B=0\\
   C=\frac{20}{243}+\frac{8(3+4\ln{2})}{729\pi^2}x^2-\frac{160\ln{3}}{243\pi^2}
x^2\quad;\quad
    L=\frac{160}{243\pi^2}x^2  \ .
\end{aligned}
\end{equation}

\section{Calculation of the second-order term}

\label{appendix:c}
In the following, we detail the calculation of the second-order term for the 
energy,

\begin{equation}
\begin{aligned}
     \frac{E}{N}=\epsilon_F\frac{1}{\nu k_F^5}\sum_{\lambda_1,\lambda_2}I_2(k_{F,\lambda_1},k_{F,\lambda_2})a_0^2(1-\delta_{\lambda_1,\lambda_2});\\
     \mbox{where}\quad I_2(k_{F,\lambda_1},k_{F,\lambda_2})=\frac{3}{16\pi^5}\int d\textbf{l}n_l\int d\textbf{k}n_k \int2d\textbf{q}d\textbf{q}'\frac{1-(1-n_q)(1-n_{q'})}{q^2+q'^2-k^2-l^2}\delta(\textbf{q}+\textbf{q}'-\textbf{k}-\textbf{l}) \ .
\end{aligned}
\end{equation}
We need to calculate the multiple integral 
$I_2$. The first step is to expand the inner part, where the 
occupation functions appear. One, then, obtains an expression with 
$n_q+n_{q'}-n_qn_{q'}$. The terms containing $n_q$ and $n_{q'}$ correspond to 
integrals of spheres (SI) with respect to $\textbf{q}$ or $\textbf{q}'$, which 
can be integrated. The rest of the integrals with respect to $\textbf{k}$ and 
$\textbf{l}$ are quite arduous to integrate, but after some lengthy calculations, 
one can obtain them. And finally, the term proportional to $n_qn_{q'}$, 
corresponding to the volume of the intersection of two spheres (SII) is 
zero due to symmetry reasons. 

We rewrite the expression of $I_2$ using 
the integrated functions SI and SII,
\begin{equation*}
   I_2(k_{F,\lambda_1},k_{F,\lambda_2})=\frac{3}{16\pi^5}\int d\textbf{l}n_l\int d\textbf{k}n_k \int2d\textbf{q}d\textbf{q}'\frac{n_q+n_{q'}-n_q n_{q'}}{q^2+q'^2-k^2-l^2}\delta(\textbf{q}+\textbf{q}'-\textbf{k}-\textbf{l})
\end{equation*}
\begin{equation}
    =\frac{3}{16\pi^5}\int d\textbf{l}n_l\int d\textbf{k}n_k\bigg[SI(P,R,k_{F,\lambda_1})+SI(P,R,k_{F,\lambda_2})-SII(P,R,k_{F,\lambda_1},k_{F,\lambda_2})\bigg]
\end{equation}

First of all, let's show that the SII term is zero. In order to follow the 
derivation, we need to have in mind that the momenta $\textbf{k}$ and $\textbf{q}$ 
run over the same momentum $k_{F,\lambda_1}$, and that the momenta $\textbf{l}$ 
and $\textbf{q}'$ do so over $k_{F,\lambda_2}$. The procedure is the following. We 
split the integral in identical parts. The first part is integrated with respect 
to $\textbf{l}$, the second one with respect to $\textbf{q}'$, both running 
over the same values. We note that it could have been done integrating with 
respect to $\textbf{k}$ and $\textbf{q}$, instead of $\textbf{l}$ and 
$\textbf{q}'$. Then, we slightly manipulate the two expressions and we obtain 
two identical integrals but with opposite signs, hence, they cancel each other,
\begin{equation}
\begin{aligned}
    \int d\textbf{l}n_l\int d\textbf{k}n_k \int2d\textbf{q}d\textbf{q}'\frac{n_qn_{q'}}{q^2+q'^2-k^2-l^2}\delta(\textbf{q}+\textbf{q}'-\textbf{k}-\textbf{l})=2\int d\textbf{l} d\textbf{k} d\textbf{q}d\textbf{q}'\frac{n_ln_kn_qn_{q'}}{q^2+q'^2-k^2-l^2}\delta(\textbf{q}+\textbf{q}'-\textbf{k}-\textbf{l})\\
    =\int d\textbf{l} d\textbf{k} d\textbf{q}d\textbf{q}'\frac{n_ln_kn_qn_{q'}}{q^2+q'^2-k^2-l^2}\delta(\textbf{q}+\textbf{q}'-\textbf{k}-\textbf{l})+\int d\textbf{l} d\textbf{k}  d\textbf{q}d\textbf{q}'\frac{n_ln_kn_qn_{q'}}{q^2+q'^2-k^2-l^2}\delta(\textbf{q}+\textbf{q}'-\textbf{k}-\textbf{l})\\
    =\int d\textbf{k} d\textbf{q}d\textbf{q}'\frac{n_kn_qn_{q'}}{-2k^2+2\textbf{k}(\textbf{q}+\textbf{q}')-2\textbf{q}\textbf{q}'}+\int d\textbf{l} d\textbf{k} d\textbf{q}\frac{n_ln_kn_q}{2q^2-2\textbf{q}(\textbf{k}+\textbf{l})+2\textbf{k}\textbf{l}}\\
    =-\frac{1}{2}\int d\textbf{k} d\textbf{q}d\textbf{q}'\frac{n_kn_qn_{q'}}{k^2-\textbf{k}(\textbf{q}+\textbf{q}')+\textbf{q}\textbf{q}'}+\frac{1}{2}\int d\textbf{q} d\textbf{k} d\textbf{l}\frac{n_qn_kn_l}{q^2-\textbf{q}(\textbf{k}+\textbf{l})+\textbf{k}\textbf{l}}=0
\end{aligned}
\end{equation}

With this, $I_2$ becomes
\begin{equation}
    I_2(k_{F,\lambda_1},k_{F,\lambda_2})=\frac{3}{16\pi^5}\int d\textbf{l}n_l\int d\textbf{k}n_k\bigg[SI(P,R,k_{F,\lambda_1})+SI(P,R,k_{F,\lambda_2})\bigg]
\end{equation}
We will integrate only one SI; the other one will be the same but interchanging 
the Fermi momenta. In the end, we will add both expressions. 
The inner part of the integral is
\begin{equation*}
    SI(k,l,k_{F,\lambda})=\int2d\textbf{q}d\textbf{q}'\frac{n_q}{q^2+q'^2-k^2-l^2}\delta(\textbf{q}+\textbf{q}'-\textbf{k}-\textbf{l})=\int d\textbf{q}\frac{n_q}{q^2-\textbf{q}\cdot(\textbf{k}+\textbf{l})+\textbf{k}\cdot\textbf{l}}
\end{equation*}
\begin{equation*}
    =\int_0^{k_{F,\lambda}} q^2dq\int_0^{\pi}\sin{\theta}d\theta\frac{2\pi}{q^2-q\vert\textbf{k}+\textbf{l}\vert+\textbf{k}\cdot\textbf{l}}=2\pi\int_0^{k_{F,\lambda}} q^2dq\frac{1}{q\vert\textbf{k}+\textbf{l}\vert}\ln{\bigg|\frac{q^2+q\vert\textbf{k}+\textbf{l}\vert+\textbf{k}\cdot\textbf{l}}{q^2-q\vert\textbf{k}+\textbf{l}\vert+\textbf{k}\cdot\textbf{l}}\bigg|}
\end{equation*}
\begin{equation*}
    =\frac{2\pi}{\vert\textbf{k}+\textbf{l}\vert}\bigg\{\bigg(\frac{k_{F,\lambda}^2}{2}-\frac{k^2+l^2}{4}\bigg)\ln{\bigg|\frac{k_{F,\lambda}^2+k_{F,\lambda}\vert\textbf{k}+\textbf{l}\vert+\textbf{k}\cdot\textbf{l}}{k_{F,\lambda}^2-k_{F,\lambda}\vert\textbf{k}+\textbf{l}\vert+\textbf{k}\cdot\textbf{l}}\bigg|}
\end{equation*}
\begin{equation}
    -\frac{\vert\textbf{k}+\textbf{l}\vert\vert\textbf{k}-\textbf{l}\vert}{4}\ln{\bigg|\frac{k_{F,\lambda}^2+k_{F,\lambda}\vert\textbf{k}-\textbf{l}\vert-\textbf{k}\cdot\textbf{l}}{k_{F,\lambda}^2-k_{F,\lambda}\vert\textbf{k}-\textbf{l}\vert-\textbf{k}\cdot\textbf{l}}\bigg|}+k_{F,\lambda}\vert\textbf{k}+\textbf{l}\vert\bigg\}
\end{equation}
We write everything in terms of the modules of $\textbf{k}$ and $\textbf{l}$ 
and the angle between them. In the angular part, we change the variable to 
$x=-\cos{\theta}$,
\begin{equation}
    \begin{aligned}
        \vert\textbf{k}+\textbf{l}\vert=\sqrt{k^2+l^2+2kl\cos{\theta}}=\sqrt{k^2+l^2-2klx}\\
        \vert\textbf{k}-\textbf{l}\vert=\sqrt{k^2+l^2-2kl\cos{\theta}}=\sqrt{k^2+l^2+2klx}\\
        \textbf{k}\cdot\textbf{l}=kl\cos{\theta}=-klx\\
        \int d\textbf{l}n_l\int d\textbf{k}n_k=2(2\pi)^2\int_0^{k_{F,\lambda_1}}k^2dk\int_0^{k_{F,\lambda_2}}l^2dl\int_0^{\pi}\sin{\theta}d\theta=2(2\pi)^2\int_0^{k_{F,\lambda_1}}k^2dk\int_0^{k_{F,\lambda_2}}l^2dl\int_{-1}^1dx
    \end{aligned}
\end{equation}
We take out the $2\pi$ from SI and integrate over x,
\begin{equation}
\begin{aligned}
    \frac{1}{2\pi}\int dxSI(k,l,q)=\frac{2}{3}qx+\frac{1}{kl}\bigg(\frac{k^2}{4}+\frac{l^2}{4}-\frac{q^2}{2}\bigg)\sqrt{k^2+l^2-2klx}\ln{\bigg|\frac{q^2+q\sqrt{k^2+l^2-2klx}-klx}{q^2-q\sqrt{k^2+l^2-2klx}-klx}\bigg|}\\ -\frac{1}{12kl}(k^2+l^2+2klx)^{3/2}\ln{\bigg|\frac{q^2+q\sqrt{k^2+l^2+2klx}+klx}{q^2-q\sqrt{k^2+l^2+2klx}+klx}\bigg|}+\frac{k^4+l^4+2k^2l^2-2k^2q^2-2l^2q^2+q^4}{3kl\sqrt{k^2+l^2-q^2}}\ln{\bigg|\frac{q\sqrt{k^2+l^2-q^2}+klx}{q\sqrt{k^2+l^2-q^2}-klx}\bigg|}\\
    -\frac{q^3}{3kl}\bigg(\ln{\big|q\sqrt{k^2+l^2-q^2}+klx\big|}+\ln{\big|q\sqrt{k^2+l^2-q^2}-klx\big|}\bigg)
\end{aligned}
\end{equation}
The first external integral is
\begin{equation}
\begin{aligned}
    SX(k,l,q)=\frac{1}{2\pi}\int_{-1}^1 dxSI(k,l,q)=\frac{4}{3}qx+\frac{(k-l)(2k^2+2l^2-kl-3q^2)}{6kl}\ln{\bigg|\frac{q^2+q(k-l)-kl}{q^2-q(k-l)-kl}\bigg|}\\ -\frac{(k+l)(2k^2+2l^2+kl-3q^2)}{6kl}\ln{\bigg|\frac{q^2+q(k+l)+kl}{q^2-q(k+l)+kl}\bigg|}+\frac{4}{6kl}(k^2+l^2-q^2)^{3/2}\ln{\bigg|\frac{q\sqrt{k^2+l^2-q^2}+kl}{q\sqrt{k^2+l^2-q^2}-kl}\bigg|}
\end{aligned}
\end{equation}
The next step is the integration over $k$ and $l$. We will proceed in two ways 
(both producing the same result): \\1) $\int_0^kdk\int_0^qdlk^2l^2SX(k,l,q)$, 
\\2) $\lim_{k\to q}\int_0^kdk\int_0^ldlk^2l^2SX(k,l,q)$.\\

\noindent 1)
\begin{equation}
\begin{aligned}
    \int dlk^2l^2SX(k,l,q)=\frac{k}{120}\bigg[4klq(-4k^2+7q^2)+44l^3klq+(16k^5-40k^3q^2+30kq^4+40k^3l^2+30kl^4-60kl^2q^2)\ln{\bigg|\frac{l-q}{l+q}\bigg|}\\
    +(16l^5+40l^3(k-q)(k+q))\ln{\bigg|\frac{k-q}{k+q}\bigg|}+16(k^2+l^2-q^2)^{5/2}\ln{\bigg|\frac{q\sqrt{k^2+l^2-q^2}+kl}{q\sqrt{k^2+l^2-q^2}-kl}\bigg|}\bigg]
\end{aligned}
\label{pal1}
\end{equation}

It can be easily checked that the function (\ref{pal1}) is zero when $l=0$. Also, 
we will need to make use of the following limit,
\begin{equation}
    \lim_{l\to q}\ln{\bigg|\frac{q\sqrt{k^2+l^2-q^2}+kl}{q\sqrt{k^2+l^2-q^2}-kl}\bigg|}=-\lim_{l\to q}\ln{\bigg|\frac{l-q}{l+q}\bigg|}-\ln{\bigg|\frac{k-q}{k+q}\bigg|}-2\ln{\bigg|\frac{k+q}{k}\bigg|}
\end{equation}
The definite integral becomes
\begin{equation}
\begin{aligned}
    \int_0^q dlk^2l^2SX(k,l,q)=\frac{k}{120}\bigg[-16k^3q^2+72kq^4+(16q^5+40q^3(k-q)(k+q)-16k^5)\ln{\bigg|\frac{k-q}{k+q}\bigg|}-32k^5\ln{\bigg|\frac{k+q}{k}\bigg|}\bigg]
\end{aligned}
\end{equation}
Finally, the integral over $k$ is
\begin{equation}
\label{iresult2}
    \int_0^kdk\int_0^q dlk^2l^2SX(k,l,q)=\frac{1}{420}\bigg[-8k^5q^2+66k^3q^4+30kq^6+(-8k^7+35k^4q^3-42k^2q^5+15q^7)\ln{\bigg|\frac{k-q}{k+q}\bigg|}-16k^7\ln{\bigg|\frac{k+q}{k}\bigg|}\bigg]
\end{equation}

\noindent 2)
\begin{equation}
\begin{aligned}
    \int_0^kdk\int_0^l dlk^2l^2SX(k,l,q)=\frac{1}{840}\bigg[-16k^5lq-16kl^5q+88k^3l^3q+44k^3lq^3+44kl^3q^3+32klq^5\\
    +(16l^7-56l^5q^2+70l^3q^4+70k^4l^3+56k^2l^5-140k^2l^3q^2)\ln{\bigg|\frac{k-q}{k+q}\bigg|}\\
    +(16k^7-56k^5q^2+70k^3q^4+70k^3l^4+56k^5l^2-140k^3l^2q^2)\ln{\bigg|\frac{l-q}{l+q}\bigg|}\\
    +16(k^2+l^2-q^2)^{7/2}\ln{\bigg|\frac{q\sqrt{k^2+l^2-q^2}+kl}{q\sqrt{k^2+l^2-q^2}-kl}\bigg|}
\end{aligned}
\end{equation}
As the expression above is symmetrical with respect to $k$ and $l$, it does not 
matter which variable we choose to perform the limit to $(k,l) \to q$ since both results 
are formally equivalent. As we want to check that this second method gives 
the same expression as the one found in method 1, we will set the limit $l$ 
going to $q$,
\begin{equation}
\label{iresult1}
\begin{aligned}
    \lim_{l\to q}\int_0^kdk\int_0^ldlk^2l^2SX(k,l,q)=\\
    \frac{1}{840}\bigg[-16k^5q^2+132k^3q^4+60kq^6+(-16k^7+70k^4q^3-84k^2q^5+30q^7)\ln{\bigg|\frac{k-q}{k+q}\bigg|}-32k^7\ln{\bigg|\frac{k+q}{k}\bigg|}\bigg]
\end{aligned}
\end{equation}
We recover indeed the same expression. 

Coming back to the integral we had at the beginning, 
\begin{equation}
   I_2(k_{F,\lambda_1},k_{F,\lambda_2})=\frac{3}{16\pi^5}\int d\textbf{l}n_l\int 
d\textbf{k}n_k\bigg[SI(P,R,k_{F,\lambda_1})+SI(P,R,k_{F,\lambda_2})\bigg]
\end{equation}
We substitute the results we have obtained for the integrals (Eqs. 
(\ref{iresult2}) or (\ref{iresult1})) and we recover the factor $2(2\pi)^3$ 
coming from the angular integrals.
\begin{equation}
\begin{aligned}
    I_2=\frac{3}{16\pi^5}2(2\pi)^3\bigg(\lim_{k\to 
q}\int_0^kdk\int_0^ldlk^2l^2SX(k,l,q)+\lim_{l\to 
q'}\int_0^kdk\int_0^ldlk^2l^2SX(k,l,q')\bigg)\\
   =\frac{3}{16\pi^5}\frac{2(2\pi)^3}{420}\bigg[
-8k^5l^2-8l^5k^2+66k^3l^4+66l^3k^4+30kl^6+30lk^6\\  
+(7k^7+7l^7+35k^4l^3+35l^4k^3-42k^2l^5-42l^2k^5)\ln{\bigg|\frac{k-l}{k+l}\bigg|}
-16k^7\ln{\bigg|\frac{k+l}{k}\bigg|}-16l^7\ln{\bigg|\frac{k+l}{l}\bigg|}\bigg]\\
    =\frac{1}{140\pi^2}\bigg[2kl(k+l)(15k^4-19k^3l+52k^2l^2-19kl^3+15l^4)\\     
+7(k+l)(k-l)^4(k^2+3kl+l^2)\ln{\bigg|\frac{k-l}{k+l}\bigg|}-16\bigg(k^7\ln{
\bigg|\frac{k+l}{k}\bigg|}+l^7\ln{\bigg|\frac{k+l}{l}\bigg|}\bigg)\bigg]\\
\end{aligned}
\end{equation}
After putting everything together and rearranging terms, $I_2$ is written in 
terms of k and l as
\begin{equation}
\begin{aligned}
   I_2=\frac{4}{35\pi^2}\bigg[\frac{1}{8}
kl(k+l)(15k^4-19k^3l+52k^2l^2-19kl^3+15l^4)\\
   +\frac{7}{16}(k+l)(k-l)^4(k^2+3kl+l^2)\ln{\bigg|\frac{k-l}{k+l}\bigg|}
-\bigg(k^7\ln{\bigg|\frac{k+l}{k}\bigg|}+l^7\ln{\bigg|\frac{k+l}{l}\bigg|}
\bigg)\bigg]
\end{aligned}
\end{equation}
Replacing $k$ and $l$ by the Fermi momenta ($k_{F,\lambda_1}$ and 
$k_{F,\lambda_2}$),
\begin{equation}
\begin{aligned}
    I_2=\frac{4}{35\pi^2}\bigg[\frac{1}{8}k_{F,\lambda_1}k_{F,\lambda_2}(k_{F,\lambda_1}+k_{F,\lambda_2})(15k_{F,\lambda_1}^4-19k_{F,\lambda_1}^3k_{F,\lambda_2}+52k_{F,\lambda_1}^2k_{F,\lambda_2}^2-19k_{F,\lambda_1}k_{F,\lambda_2}^3+15k_{F,\lambda_2}^4)\\
    +\frac{7}{16}(k_{F,\lambda_1}+k_{F,\lambda_2})(k_{F,\lambda_1}-k_{F,\lambda_2})^4(k_{F,\lambda_1}^2+3k_{F,\lambda_1}k_{F,\lambda_2}+k_{F,\lambda_2}^2)\ln{\bigg|\frac{k_{F,\lambda_1}-k_{F,\lambda_2}}{k_{F,\lambda_1}+k_{F,\lambda_2}}\bigg|}\\
    -\bigg(k_{F,\lambda_1}^7\ln{\bigg|\frac{k_{F,\lambda_1}+k_{F,\lambda_2}}{k_{F,\lambda_1}}\bigg|}+k_{F,\lambda_2}^7\ln{\bigg|\frac{k_{F,\lambda_1}+k_{F,\lambda_2}}{k_{F,\lambda_2}}\bigg|}\bigg)\bigg]
\end{aligned}
\end{equation}
Now, we replace the Fermi momenta by $k_FC_{\lambda}^{1/3}$,
\begin{equation}
\begin{aligned}
    I_2=\frac{4k_F^7}{35\pi^2}\bigg[\frac{1}{8}C_{\lambda_1}^{1/3}C_{\lambda_2}^{1/3}\big(C_{\lambda_1}^{1/3}+C_{\lambda_2}^{1/3}\big)\big(15C_{\lambda_1}^{4/3}-19C_{\lambda_1}C_{\lambda_2}^{1/3}+52C_{\lambda_1}^{2/3}C_{\lambda_2}^{2/3}-19C_{\lambda_1}^{1/3}C_{\lambda_2}+15C_{\lambda_2}^{4/3}\big)\\
    +\frac{7}{16}\big(C_{\lambda_1}^{1/3}+C_{\lambda_2}^{1/3}\big)\big(C_{\lambda_1}^{1/3}-C_{\lambda_2}^{1/3}\big)^4\big(C_{\lambda_1}^{2/3}+3C_{\lambda_1}^{1/3}C_{\lambda_2}^{1/3}+C_{\lambda_2}^{2/3}\big)\ln{\bigg|\frac{C_{\lambda_1}^{1/3}-C_{\lambda_2}^{1/3}}{C_{\lambda_1}^{1/3}+C_{\lambda_2}^{1/3}}\bigg|}\\
    -\bigg(C_{\lambda_1}^{7/3}\ln{\bigg|\frac{C_{\lambda_1}^{1/3}+C_{\lambda_2}^{1/3}}{C_{\lambda_1}^{1/3}}\bigg|}+C_{\lambda_2}^{7/3}\ln{\bigg|\frac{C_{\lambda_1}^{1/3}+C_{\lambda_2}^{1/3}}{C_{\lambda_2}^{1/3}}\bigg|}\bigg)\bigg]
\end{aligned}
\end{equation}
Finally, in terms of $y=(C_{\lambda_1}/C_{\lambda_2})^{1/3}$,
\begin{equation}
\begin{aligned}
I_2=\frac{4k_F^7}{35\pi^2}C_{\lambda_1}C_{\lambda_2}\frac{C_{\lambda_1}^{1/3}+C_
{\lambda_2}^{1/3}}{2}\bigg[\frac{1}{4}\big(15 y^2-19 y+52-19y^{-1}+15y^{-2}\big)
+\frac{7}{8} 
y^{-2}\big(y-1\big)^4\big(y+3+y^{-1}\big)\ln{\bigg\vert\frac{1-y}{1+y 
}\bigg\vert}\\
    -\frac{2y^4}{1+y}\ln{\bigg\vert 
1+y^{-1}\bigg\vert}-\frac{2y^{-4}}{1+y^{-1}}\ln{\bigg\vert 1+y\bigg\vert}\bigg]
\end{aligned}
\end{equation}

\end{widetext}

\bibliography{refs}

\end{document}